\documentclass[aps,prl,showpacs,preprint,floatfix]{revtex4}
\usepackage{latexsym}
\usepackage{amsfonts}
\usepackage{amssymb}
\usepackage{amsmath}
\usepackage{graphicx}

\begin{document}

\title{Coupling efficiency for phase locking of a spin transfer oscillator to a microwave current}

\author{B. Georges, J. Grollier, M. Darques,  V. Cros, C. Deranlot, B. Marcilhac, A. Fert}
\affiliation{Unit\'e Mixte de Physique CNRS/Thales and Universit\'e
Paris Sud 11, RD 128, 91767 Palaiseau, France}
\author{G. Faini}
\affiliation{CNRS, Phynano team, Laboratoire de Photonique et de Nanostructures, Route de Nozay, 91400 Marcoussis, France}

\begin{abstract}
The phase locking behavior of spin transfer nano-oscillators (STNOs) to an external microwave signal is experimentally studied as a function of the STNO intrinsic parameters. We extract the coupling strength from our data using the derived phase dynamics of a forced STNO. The predicted trends on the coupling strength for phase locking as a function of intrinsic features of the oscillators i.e. power, linewidth, agility in current, are central to optimize the emitted power in arrays of mutually coupled STNOs.
\end{abstract}

\pacs{85.75.-d,75.47.-m,75.40.Gb}\maketitle

The prediction by Slonczewski \cite{Slonc} and Berger \cite{Berger} of magnetization dynamics induced by a spin polarized current has stimulated many experimental and theoretical studies in the last decade. Among them, a strong interest of spin transfer effects is the observation of a steady precession of a magnetization, generating a microwave power due to the magnetoresistive response of the devices \cite{KiselevRippard,RippardPRL,TsoiNature}. Numerous efforts were made to understand the microwave emission under magnetic field and current bias in a single spin transfer nano-oscillator (STNO) \cite{SlavinIEEE,KrivoBerkov,PufallVortex,KimMistral,BoulleNature,Bertotti}. Most of the characteristics presently observed in STNOs are indeed very attractive for future applications in telecommunication devices. Their frequency window is in the GHz range and their agility is huge compared to standard voltage or current controlled oscillators. Moreover STNOs are fully compatible with a high level of integration. However a breakthrough has to be achieved regarding the output power of a single STNO that remains dramatically too weak (typically in the low 1nW range). Thus, a major challenge is to synchronize many of these STNOs in order to increase the microwave power and reduce the linewidth of the emission\cite{Tsoi,SlavinPRBsource,SlavinPRBmutual,Akerman,Rezende,Zhang}. The coupling between several STNOs may result from various local or non local mechanisms. For example, the coupling between two closely spaced STNOs via the interaction of spin-waves propagating into a common magnetic layer is only efficient over the spin wave decay length i.e around one micrometer. Recent experiments in a point contact geometry have demonstrated the synchronization of two oscillators by this local coupling \cite{Kaka,Mancoff}. On the contrary, a long range coupling between STNOs could exist under the interaction through spin transfer self-emitted microwave currents. We have recently predicted that this mechanism, based on the interaction of electrically connected STNOs through their common emitted microwave current, can be strong enough to achieve a mutual phase locking between several STNOs \cite{Grollier}.

In order to investigate this latter coupling mechanism, we have performed injection locking experiments of a single STNO to an external microwave signal. Recently, 
W.H. Rippard \textit{et al.} \cite{RippardPRLinjection} have already shown that a STNO nanocontact can be locked to an external signal. In the present letter, we carry through a detailed study on the influence of some intrinsic emission characteristics of a STNO (linewidth, agility in current) on the coupling strength to the external microwave current $I_{hf}$. These intrinsic parameters are tunable almost independently by choosing appropriate values of the injected dc current $I_{dc}$ and applied field. The measured variations of the coupling strength on the STNOs characteristics are analysed within the approximation of weakly forced non linear oscillators. The identification of some clear trends for the coupling strength of coupled STNOs is an important step on the path to the synchronization of large numbers of STNOs in arrays.

The present experiments have been performed at room temperature on  a 70 x 170 nm$^2$ elliptic nanopillar patterned from a Py(15nm)/Cu(10nm)/Py(3nm) spin valve (Py=Ni$_{80}$Fe$_{20}$). The device resistance is 24.5 $\Omega$ and its magnetoresistance variation is 82 m$\Omega$. In our convention, a positive current is defined as electrons flowing from the thick to the thin magnetic layer. For high frequency measurements, a  dc current ranging from -5 to -8 mA is injected in the nanopillar through a bias tee under an external magnetic field H$_\bot$ applied perpendicular to the layers, ranging from 2 to 3 kOe. The emitted power is detected using a spectrum analyzer after a + 35 dB amplification. In addition, a microwave circulator is introduced in the circuit between the bias tee and the analyzer. This device allows the injection of large microwave currents $I_{hf}$ from a source. The microwave reflection coefficient has been evaluated using a network analyzer and is almost constant (about 50 $\%$) within the working frequency range (1 to 3 GHz). 

\begin{figure}[h]
   \centering
    \includegraphics[width=0.5\textwidth]{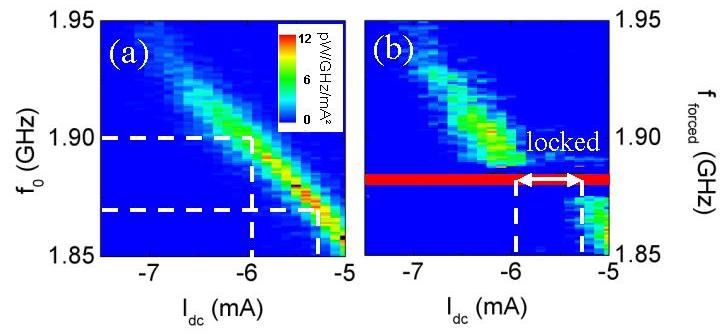} 
     \caption{Georges et al.}
\label{fig1}
\end{figure}

In Fig.\ref{fig1}(a), we display a map of the microwave emission power (in linear  color scale) versus frequency and dc current under a constant applied field of 2.65 kOe. The STNOs emission characteristics i.e. the free running frequency f$_0$ (free running meaning without external microwave signal), the power amplitude, the linewidth and the local agility in current depend on the current $I_{dc}$ at fixed applied fields. A similar characterization in presence of an external microwave signal of - 22 dBm at f$_{source}$ = 1.88 GHz is presented in Fig.\ref{fig1}(b).

Two regimes can be defined from Fig.\ref{fig1}(b). The first one, called the locking regime, exists for $I_{dc}$ between -5.3 mA and -5.9 mA in which the STNO frequency, called  f$_{forced}$, is locked to the source frequency. The measured locking range in current (0.6 mA) is converted in frequency unit (20 MHz) by using the frequency variation in current of the free running oscillator f$_0$ as shown in Fig.\ref{fig1}(a) by the dotted white lines. Outside this current window, the STNO is no longer beating at the source frequency but deviates also from its natural frequency f$_0$. This behavior is commonly defined as the pulling regime and it is characterized by the frequency shift $\Delta$f = f$_{forced}$ - f$_0$. 

\begin{figure}[h]
   \centering
    \includegraphics[width=0.5\textwidth]{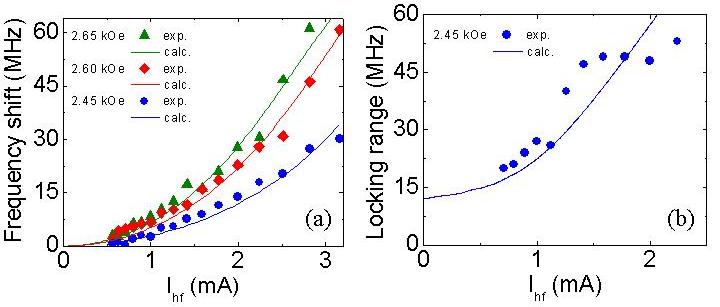} 
     \caption{Georges et al.} 
    \label{fig2}
\end{figure}

Experimentally, the frequency shift measurement is more accurate than the locking range measurement because close to the source frequency, the signal of the forced oscillator is partly hidden by the external microwave signal. In Fig.\ref{fig2}(a), the experimental frequency shift $\Delta$f as a function of $I_{hf}$ measured for H$_\bot$ = 2.45, 2.60 and 2.65 kOe, I$_{dc}$=-6 mA and f$_{source}$=1.80 GHz is presented. By changing the applied field H$_\bot$, the characteristics of the free running oscillation differ mainly by its agility in current df$_{0}$/dI$_{dc}$. We measure 37, 52 and 72 MHz/mA respectively for 2.45, 2.60 and 2.65 kOe. As can be seen from Fig.\ref{fig2}(a), at a fixed microwave current $I_{hf}$, the frequency shift $\Delta$f increases with the agility in current. The experimental locking range is plotted in Fig.\ref{fig2}(b) as a function of the microwave current $I_{hf}$ for H$_\bot$=2.45 kOe and f$_{source}$=1.88 GHz.  Similarly to the frequency shift, the locking range increases with $I_{hf}$. This trend has already been observed by Rippard \textit{et al.} \cite{RippardPRLinjection}. From the evolution of the locking range and the frequency shift with $I_{hf}$, we clearly see that the emission characteristics of the STNO modify its coupling to the external signal.

A quantitative analysis of the coupling strength as a function of the emission characteristics requires to formulate the theory of weakly forced oscillators to our case of interest i.e. STNOs. From the equation of magnetization motion, we derive the equation for the phase dynamics of the spin transfer oscillator in presence of an external signal. We start from the equation for the dimensionless complex amplitude of the spin wave mode b = c.e$^{j\varphi}$ ($\varphi$ is the phase of the oscillation) recently derived by A. Slavin \textit{et al.} \cite{SlavinIEEE} in which we neglect for simplicity the non linear damping parameter, Q = 0: 
\begin{eqnarray}\label{Slavin}
\frac{db}{dt} & = & -i(\omega_{FMR} + N b^2)b - \Gamma b + \sigma I_{dc} (1 - b^2) \nonumber \\
& + & \frac{\sigma tan(\gamma)}{2\sqrt{2}} I_{hf}e^{-i \omega_{source}t}
\end{eqnarray} 
where $\omega_{FMR}$ is the resonance frequency in the absence of the spin transfer torque, N the non-linear frequency shift, $\Gamma$ the Gilbert damping, $\sigma$ the spin transfer efficiency and $I_{dc}$ the dc current. The last term on the right side describes the external microwave signal with $\gamma$ being the equilibrium angle between the fixed and free magnetization. For I$_{hf}$=0 (no source), we obtain from Eq.(\ref{Slavin}) the expression of the uniform rotating phase $\Phi$ that is valid even when the oscillator is slightly perturbed out of its limit cycle: 
\begin{equation}
\Phi=\varphi+\frac{N}{\sigma I_{dc}} ln(c)+ \Phi_0
\end{equation}
Now considering the case of small perturbations to the limit cycle, we can derive the phase dynamics from Eq.(\ref{Slavin}) resulting in the following general equation of the phase dynamics of a forced non linear oscillator\cite{Adler} :
\begin{equation}\label{Adler}
\frac{d(\Delta\Phi)}{dt} = - 2 \pi \Delta f_{det} + \epsilon sin(\Delta\Phi) + \xi(t)
\end{equation} 
where $\Delta$$\Phi$ is the phase difference between the oscillator and the external microwave signal. The detuning $\Delta f_{det}$ is defined as f$_0$ - f$_{source}$. The parameter $\epsilon$ characterizes the coupling strength between the oscillator and the external microwave signal. The third term $\xi(t)$ expresses the Gaussian white noise that accounts for a linewidth $w^{2}$ of the spectral peak. We can now express the coupling strength $\epsilon$ for the case of STNOs as a function of experimentally available parameters : the dc current threshold $I_{th}$ for the onset of the oscillations is expressed as $I_{th} = \Gamma / \sigma$, and the non linear frequency shift N = $\frac{d\omega}{dc^2}$ is $2\pi \frac{\partial f_0}{\partial I_{dc}}\frac{I_{dc}^2}{I_{th}}$. The ratio $\frac{\partial f_0}{\partial I_{dc}}$ is the agility in current. We thus obtain:
\begin{equation}\label{epsilonbis}
\epsilon = \sigma tan(\gamma) \frac{I_{hf}}{2\sqrt{2}} \sqrt{\frac{I_{dc}}{I_{dc} - I_{th}}} \sqrt{1 + \left(\frac{2 \pi I_{dc}}{\sigma I_{th}} \frac{\partial f_0}{\partial I_{dc}}\right)^2}
\end{equation}
Note that for $\xi(t)=0$, the coupling strength $\epsilon$ is the equivalent of the locking range. For this particular case, our calculation of the noiseless locking range is similar to the locking range derived using another method by Slavin \textit{et al.} (see Eq.(16) in Ref.\cite{SlavinPRBsource}). 
Taking into account the noise ($\xi(t)\neq0$), we derive from Eq.(\ref{Adler})\cite{Pikovski} an analytical expression of the frequency mismatch $\Delta f_{mis}$ defined as f$_{forced}$ - f$_{source}$ : 
\begin{eqnarray}
\Delta f_{mis} = f_{forced}-f_{source} = - \Delta f_{det} -
\nonumber \\
- \epsilon Im \left\{\frac{1}{2\frac{i\Delta f_{det}-w^2}{\epsilon}+\frac{1}{2\frac{i\Delta f_{det} - 2 w^2}{\epsilon}+\frac{1}{2\frac{i\Delta f_{det} -3 w^2}{\epsilon}+...}}}\right\}
\label{mismatch}
\end{eqnarray}

\begin{figure}[h]
   \centering
    \includegraphics[width=0.5\textwidth]{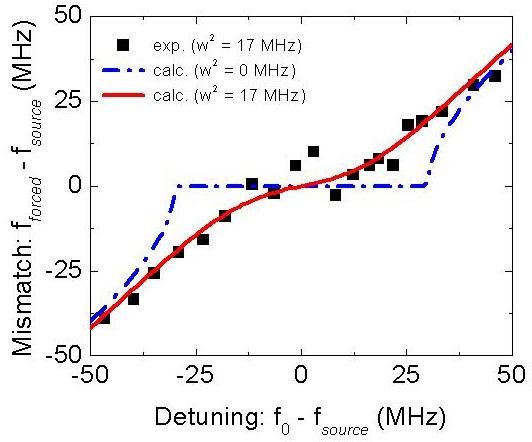} 
     \caption{Georges et al.}
    \label{fig3}
\end{figure}

In Fig.\ref{fig3}, we display the experimental mismatch $f_{forced}-f_{source}$ versus the detuning $f_{0}-f_{source}$ (black dots). These measurements are done with H$_\bot$ = 2.60 kOe, $I_{hf}$=1.1 mA at $f_{source}$=1.90 GHz. The frequency of the STNO is changed by varying the dc current $I_{dc}$ from -5 to -8 mA : in this window, both the linewidth and the agility are constant. We emphasize that it is a necessary condition to compare with the model. To obtain the blue dotted line in Fig.\ref{fig3}, we calculate the mismatch from Eq.(\ref{mismatch}) without noise ($\xi(t)$ = 0), the coupling strength $\epsilon$ being the only free parameter. There is a large disagreement with the experimental data especially close to zero mismatch where the phase locking takes place. Now we integrate in the calculation the actual linewidth (17 MHz) to account for the noise. The overall agreement is excellent (red plain line in Fig.\ref{fig3}) and yields a reliable value of 50 MHz for the coupling strength $\epsilon$. This proves the important role played by the noise that induces fluctuations in the phase dynamics of STNOs and therefore weakens the synchronization as predicted by Adler's model\cite{Pikovski}. Moreover it validates our choice of the classical model of forced non linear oscillators to describe the phase locking in STNOs. 

Thus we are able to determine using Eq.(\ref{mismatch}) the ratio $\epsilon$/$I_{hf}$ which , in turn, allows us to calculate the frequency shift and the locking range as a function of $I_{hf}$. These calculations are plotted as plain lines in Fig.\ref{fig2} (a) and (b). The agreement with the experimental data is excellent, in particular for the frequency shift vs  $I_{hf}$. For large values of the microwave current, the evolution of the locking range is linear with $I_{hf}$. On the contrary, when the locking range becomes comparable to the linewidth, the noise becomes predominant. 

\begin{figure}[h]
   \centering
    \includegraphics[width=0.5\textwidth]{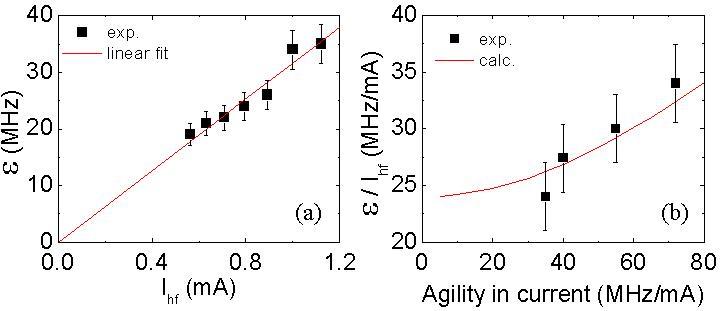} 
     \caption{Georges et al.}
    \label{fig4}
\end{figure}

The next step is to test experimentally the derived expression of $\epsilon$ in Eq.(\ref{epsilonbis}). For this purpose, we have repeated the  measurement depicted in Fig.\ref{fig3} for several values of the microwave current. For each $I_{hf}$, we obtain a value of the coupling strength $\epsilon$ from the best fit using Eq.(\ref{mismatch}). The resulting dependence of $\epsilon$ as a function of $I_{hf}$ is shown in Fig.\ref{fig4}(a). In agreement with Eq.(\ref{epsilonbis}), the coupling strength $\epsilon$ is proportional to the injected microwave current. A linear fit to the data yields a slope of 30 MHz/mA. We have also studied the influence of the agility in current on the coupling efficiency. We have chosen experimental conditions for which the dc current $I_{dc}$, the threshold current $I_{th}$ and the peak linewidth remain constant but the agility changes. The resulting evolution of $\epsilon$/$I_{hf}$ as a function of the agility in current is plotted as black squares in Fig.\ref{fig4}(b). In addition, we have calculated the same evolution using Eq.(\ref{epsilonbis}) taking $\gamma$ = 2.75$^{\circ}$ and $\sigma$ = 1 GHz/mA (red plain line in Fig.\ref{fig4}(b)). The value of the equilibrium angle $\gamma$ is reasonable considering the experimental conditions and the value of $\sigma$ leads to a reliable spin polarization coefficient (0.31). As predicted by the model, the more agile the oscillator is, the more efficient is the coupling to the external signal. 

To conclude, we have shown experimentally the influence of the linewidth and agility in current on the coupling between a nanopillar spin transfer oscillator and an external microwave signal. We have derived analytically the phase dynamics of a forced STNO taking into account the noise contribution. The derived expression of the coupling strength depends only on experimentally available parameters. We have successfully tested the predictions of our calculation i.e. the coupling strength depends on the agility in current and is proportional to the microwave current $I_{hf}$ provided the source. Typical values of $I_{hf}$ needed to phase lock a single STNO are about 1 mA. This provides an estimation of the total microwave currents that should be emitted by an STNO assembly in order to be synchronized.

This work was partly supported by the French National Agency of Research ANR through the PNANO program (NANOMASER PNANO-06-067-04) and the EU network SPINSWITCH (MRTN-CT-2006-035327).

\newpage

\textbf{Figure captions}\\

Figure 1. Map of the emission amplitude (in linear color scale) as a function of the frequency and the injected dc current with an out of plane magnetic field of 2.65 kOe. (a) free running emission (b) emission with a -22 dBm injected external microwave signal at 1.88 GHz. \\

Figure 2. (a) Frequency shift, f$_{\textit{forced}}$ - f$_{0}$, as a function of the microwave current  $I_{hf}$ with an external signal f$_{source}$ set at 1.80 GHz. The experimental data have been obtained at H$_\bot$ = 2.45, 2.60 and 2.65 kOe for a current $I_{dc}$ = -6 mA.  For these parameters, the free running frequency f$_{0}$ is 1.878 GHz and the linewidth is 15 MHz. 
	(b) Experimental locking range as a function of the microwave current $I_{hf}$. The measurements have been repeated for several fields i.e. 2.45, 2.60 and 2.65 kOe. The frequency of the external signal has been kept fixed at f$_{source}$ = 1.88 GHz while the free running frequency  f$_{0}$ is swept by varying the current $I_{dc}$ from - 5 mA to - 8 mA. Plain lines are linear fits to the experimental data.
      \\

Figure 3. Variation of the frequency mismatch $\Delta$f$_{mis}$ as a function of the frequency detuning $\Delta$f$_{det}$: the black dots are the experimental data obtained at H$_\bot$ = 2.60 kOe and for $I_{dc}$ ranging from -5 to -8 mA. In these conditions,the free running frequency f$_0$ varies from 1.84 to 1.96 GHz. The frequency of the source f$_{source}$ is set to 1.90 GHz and $I_{hf}$ = 1.1 mA. Blue and red curves are simulations according to Adler's model respectively with $w$$^{2}$ = 0 and $w$$^{2}$ = 17 MHz. The coupling strength $\epsilon$ is the only free parameter for the fit to the data. For this case, we find $\epsilon$ = 50 MHz. \\

Figure 4. (a) Variation of the coupling strength $\epsilon$ versus the microwave current $I_{hf}$ for H$_\bot$ = 2.65 kOe, $I_{th}$ = -3 mA, $I_{dc}$ = -5 to -8 mA, $\frac{\partial f_0}{\partial I_{dc}}$ = 72 MHz/mA, $w$$^{2}$ = 17 MHz . The experimental points (black dots) are the best fits to the curve mismatch vs detuning for different values of $I_{hf}$, using Eq(\ref{mismatch}) with noise corresponding to the experimental linewidth of 17 MHz (b) Black squares: experimental variation of $\epsilon$/$I_{hf}$ as a function of the agility in current for our STNO such as $I_{th}$ = -3 mA, $I_{dc}$ = -6 mA, $w$$^{2}$ = 13 MHz. Red line: calculations using Eq.(\ref{epsilonbis}) taking $\sigma$ = 1 GHz/mA and $\gamma$ = 2.75$^{\circ}$ \\

\newpage

\newpage

\newpage

\newpage

\end{document}